\documentclass[submission,copyright,creativecommons]{eptcs}
\providecommand{\event}{WACA 2025} %

\usepackage{iftex}

\usepackage[
  disable,     %
  textwidth=\marginparwidth,
  colorinlistoftodos,
  prependcaption,
  linecolor=green,backgroundcolor=green!25,bordercolor=green
  ]{todonotes}
\makeatletter
\if@todonotes@disabled
\else %
\newlength{\increase}
\paperwidth=\dimexpr \paperwidth + \increase\relax
\oddsidemargin=\dimexpr\oddsidemargin + .6\increase\relax
\evensidemargin=\dimexpr\evensidemargin + .45\increase\relax
\marginparwidth=\dimexpr \marginparwidth + .5\increase\relax
\let\tmptitle\title
\renewcommand{\title}[1]{\tmptitle{#1{\tiny\normalfont \makebox[10em]{\colorbox{red!20}{disable todos}}}}}
\fi
\makeatother

\usepackage{float}
\newfloat{lstfloat}{htbp}{lop}
\floatname{lstfloat}{Listing}
 
\usepackage{listings}
\usepackage{inconsolata}
\usepackage[capitalise,noabbrev]{cleveref}

\definecolor{color:keyword}{RGB}{150,0,150}
\definecolor{color:comment}{RGB}{0,128,0}
\definecolor{color:string}{RGB}{0,0,255}

\lstdefinelanguage{MyChor}{
morekeywords={
  while, if, else, scope, true, false, not, getInput, and, 
  prop, from, to, via, on, do, and, roles, null, rule, N, 
  aioc, preamble, starter, include, or, E, with, newRoles
},
sensitive=true,
morecomment=[l]{//},
morecomment=[s]{/*}{*/},
morestring=[b]",
otherkeywords={;,|,@,!}
}

\ifpdf
  \usepackage{underscore}         %
  \usepackage[T1]{fontenc}        %
\else
  \usepackage{breakurl}           %
\fi

\providecommand{\thisvolume}[1]{this volume of EPTCS, Open Publishing Association}

\title{Adaptable TeaStore: A Choreographic Approach\thanks{Work
    partially supported by French ANR project SmartCloud
    ANR-23-CE25-0012, by PRIN project FREEDA (CUP: I53D23003550006)
    funded by the frameworks PRIN (MUR, Italy) and Next Generation EU, by project PNRR CN HPC - SPOKE 9 - Innovation Grant LEONARDO - TASI - RTMER funded by the NextGenerationEU European initiative through the MUR,
    and by INdAM - GNCS 2024 project MARVEL, code CUP E53C23001670001}}
\author{ Giuseppe De Palma \qquad
  Saverio Giallorenzo \qquad
  Ivan Lanese \qquad
  Gianluigi Zavattaro
  \institute{Universit\`a di Bologna and INRIA\\ Bologna, Italy}
  \email{\{giuseppe.depalma2,saverio.giallorenzo,ivan.lanese,gianluigi.zavattaro\}@unibo.it}
} \def\titlerunning{Adaptable TeaStore: A Choreographic Approach}
\def\authorrunning{G. De Palma et al.}
\begin{document}

\maketitle

\begin{abstract}
The Adaptable TeaStore has recently been proposed as a reference model for adaptable microservice architectures. It includes different configurations, as well as scenarios requiring to transition between them. We describe an implementation of the Adaptable TeaStore based on AIOCJ, a choreographic language that allows one to program multiparty systems that can adapt at runtime to different conditions. Following the choreographic tradition, AIOCJ ensures by-construction correctness of communications (e.g., no deadlocks) before, during, and after adaptation. Adaptation is dynamic, and the adaptation scenarios need to be fully specified only at runtime. Using AIOCJ to model the Adaptable TeaStore, we showcase the strengths of the approach and its current limitations, providing suggestions for future directions for refining the paradigm (and the AIOCJ language, in particular), to better align it with real-world Cloud architectures.
\end{abstract}

\section{Introduction}
The Adaptable TeaStore has been recently proposed~\cite{BDGLZZ25} as a
reference model for adaptable microservice architectures. It extends the
TeaStore reference model~\cite{BDGLZZ25} for static microservice architectures
with multiple configurations as well as many adaptation scenarios that trigger
transitions between them.

In this paper, we model the Adaptable TeaStore specification as adaptable
choreographies, using AIOCJ (Adaptable Interaction-Oriented Choreographies in
Jolie)~\cite{SLE2014,LMCS2017-dynamicChoreo,book-betty}, an executable language
that allows one to program adaptable multiparty systems. 

Following the tradition of choreographic specifications and
programming~\cite{survey,HYC16,M23}, in AIOCJ, a single artefact defines a
multiparty system and one can derive the code of each component that participate
in that system from the artefact via a projection operation. More precisely, the
idea behind AIOCJ is to structure the system as a composition of
\emph{participants} and, possibly, \emph{external services}. Participants can
perform complex communication patterns and are subject to adaptation. Moreover,
participants can invoke external services, which can perform computation
external to the choreography (e.g., accessing a database). Invocations to these
services must terminate and return a value back to the invoker (a participant).
External services are not defined in AIOCJ; they can be written in any language
as long as they support invocations through one of the communication protocols
supported by AIOCJ, such as SOAP over sockets.
At adaptation, participants would consistently execute new code (as in, ``not
present in the original choreography'') that implements the adapted behaviour,
possibly reconfiguring the connections among them
and the services they have access to and changing the ways they use them.

The choreographic approach ensures by construction relevant correctness
properties of communication, such as deadlock freedom. In the specific case of
AIOCJ, these properties hold before, during, and after adaptation. As mentioned,
a distinctive trait of AIOCJ programs is that they can adapt to behaviour
unexpected at the time of writing the choreography. Concretely, when one
programs an adaptable choreography, they need to specify which parts of the code
may change in the future. The new code replacing the adaptable parts, and the
definition of the conditions triggering the adaptation, can be added afterwards
--- in particular, while the original system is running. Adaptation works by
specifying adaptation rules, which can be applied depending on the state of the
system and of its execution context/environment. A single adaptation rule may
change the code of multiple participants in a coordinated way.

Since we show an application of AIOCJ on the Adaptable TeaStore and our focus is
on the modelling of the architecture of the case study (and not, e.g., its
computational part), we use minimal implementations of the needed services, seen
(in AIOCJ) as external services.\footnote{While, in principle, one could
implement a whole architecture in AIOCJ, to separate concerns, we follow a style
that implements the coordination among components in AIOCJ and uses
services for the components' business logic.} In doing so, we follow a
programming style~\cite{GLR18} where AIOCJ participants are
adaptable connectors which coordinate the external services, %
complemented by adaptation rules allowing such coordination to change depending
on %
changing needs.

\section{Adaptable TeaStore: An Overview}
\begin{figure}[h]
    \centering
   \includegraphics[width=0.8\textwidth]{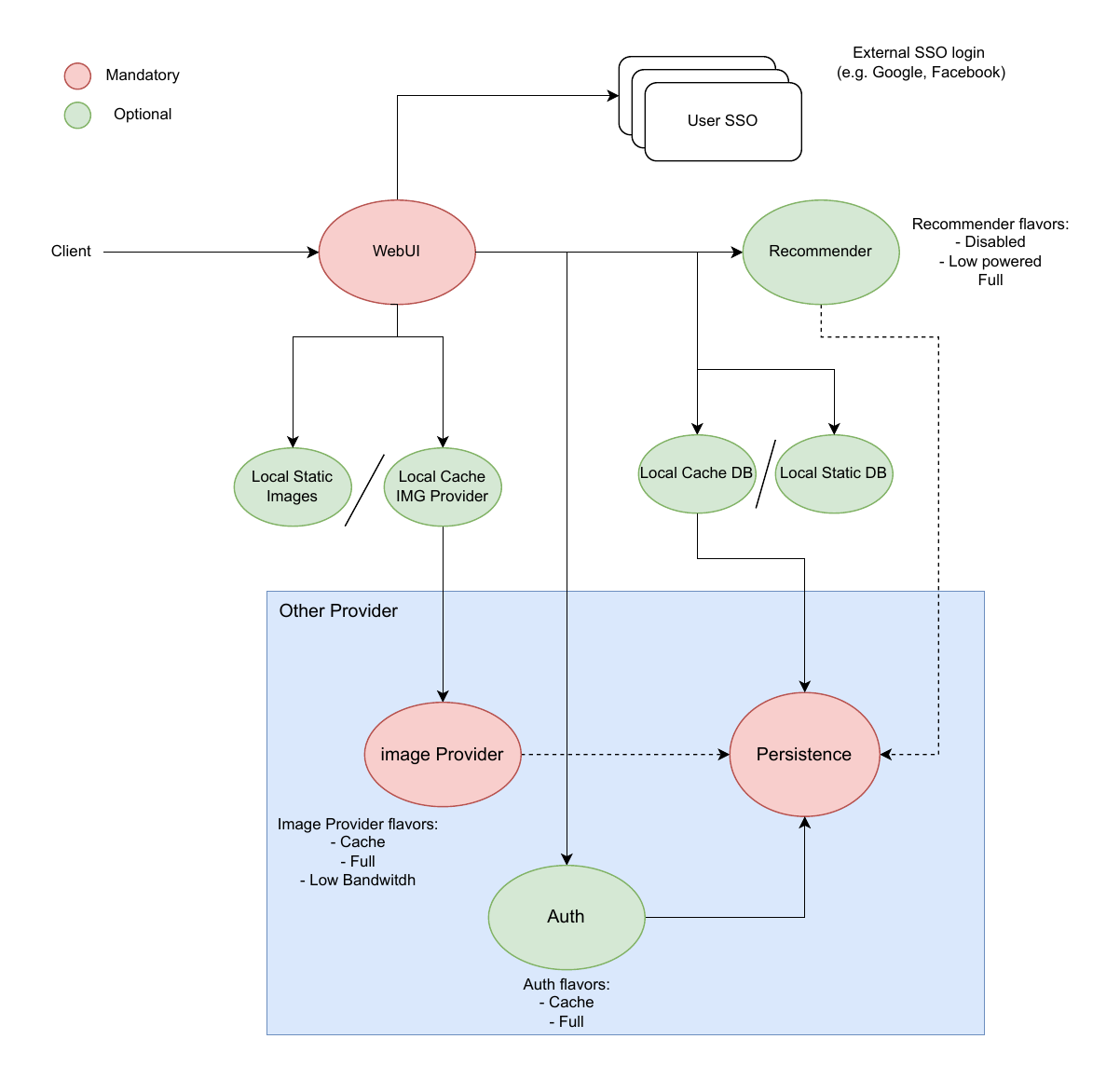}
    \caption{Adaptable TeaStore Architecture Diagram.}
    \label{fig:teastore-variant}
\end{figure}

While we refer the reader to the official Adaptable TeaStore
specification~\cite{BDGLZZ25} for full details, in this section, we briefly
summarise the Adaptable TeaStore architecture to illustrate the components at
play and the deployment modalities we choose for our AIOCJ implementation. We
report a simplified schema of the Adaptable TeaStore's architecture from the
specification~\cite{BDGLZZ25} in \cref{fig:teastore-variant}.

The Adaptable TeaStore's architecture comprises 5 main services: WebUI, Auth,
Persistence, Image Provider, and Recommender. The WebUI service works as the
entry point for users, and it orchestrates all other services. The Persistence
service acts as a layer on top of the database. We deploy it on a provider
different from the local one, making it accessible via a Local Cache. When
unavailable, a Local Static Database (with limited functionalities) replaces it.
The WebUI uses the Persistence Service to retrieve and store data, while the
Auth service uses it to retrieve user data, which it then passes to the WebUI. 

Both the Image Provider and Recommender connect to the Persistence service.
However, they only require this connection on startup (dashed lines).
Since
this startup synchronisation is not relevant for the adaptation scenarios, which are the main focus of the paper, we omit this part in our implementation.
The Image
Provider must generate an image for each product and, like Persistence, operates
remotely and connects via a Local Cache. Local Static Images can replace the
Image Provider when needed. Auth provides authentication facilities and can rely
on external User SSO services to enable authentication via, e.g., Google or
Facebook --- similarly to the startup synchronisation routine, we omit to model
SSO interactions in our implementation to focus on the main behaviour of the
application. 

The Recommender offers users suggestions about potentially interesting products.
The Recommender service is not an essential service, and can be completely
disabled.
The full-power version of the Recommender uses machine learning techniques
associated with specific user preferences, but a low-powered version based on
generic item popularity can replace it. Similarly, the Image Provider and the
Auth service can also be provided in different modalities.

\section{``Barebone'' TeaStore, in AIOCJ}

We start by presenting an AIOCJ implementation of the most basic version of
(Adaptable) TeaStore, which we call ``Barebone'' TeaStore. The Barebone version
only includes the essential services and interactions that offer minimum
functionality to users, i.e., those provided by the WebUI, the Local Static
Images, and the Local Static Persistence services. In this section, we start by
introducing a non-adaptable version of the Barebone TeaStore and, in the
subsequent sections, we proceed to present and refine the implementation with
the necessary AIOCJ constructs to make our program adaptable and able to cover
most of the Adaptable TeaStore architectural specification.

\begin{lstfloat}[t]
\begin{lstlisting}[language=MyChor,basicstyle=\ttfamily\footnotesize,caption={AIOCJ Choreography of the Barebone TeaStore version.},label={lst:bb}]
$\label{lst:bb-includes-start}$include getPageInfo from "socket://localhost:8001" with "soap"
include getPageImg from "socket://localhost:8002" with "soap"
$\label{lst:bb-includes-end}$include compilePage from "socket://localhost:8000" with "soap"

$\label{lst:bb-pre}$preamble { starter: W }

$\label{lst:bbstart}$aioc {
 address@U = getInput( "Insert address" ); 
 getPage: U( address ) -> W( address );
 $\label{lst:bb-par-start}${
  $\label{lst:bb-per-start}${
 		getPageInfo: W( address ) -> P( address );
 		info@P = getPageInfo( pid );
		getInfo: P( info ) -> W( info )
	}$\label{lst:bb-per-end}$ 
	| 
	$\label{lst:bb-img-start}${
	 	getPageImg: W( address ) -> I( address );
		img@I = getPageImg( address );
		getImg: I( img ) -> W( img )
	}$\label{lst:bb-img-end}$
 }$\label{lst:bb-par-end}$;
 $\label{lst:bb-compile-page}$page@W = compilePage( info, img );
 getPage: W( page ) -> U( page )
}$\label{lst:bbend}$
\end{lstlisting}
\end{lstfloat}

We report in \cref{lst:bb} the AIOCJ code of the Barebone TeaStore, focussing on
the interesting part at lines \ref{lst:bbstart}--\ref{lst:bbend}, contained
within the \lstinline{aioc} scope, delimited by curly brackets. Therein, we find
a choreography specifying the global behaviour of four participants: the user,
marked \lstinline{U} (acting as a client), the WebUI \lstinline{W}, the
Persistence service \lstinline{P}, and the Images service \lstinline{I}. The
choreography represents user interactions using the \lstinline{getInput}
functionality, which presents the user with a prompt through which they can
insert data. Specifically, the first action in the choreography (line 8)
concerns the \lstinline{U}ser providing the address (in this prototype the user
inserts directly the address value, while in a real application a web interface
would deal with the provision of this kind of data) of the page of the TeaStore
they want to visit; the data inserted by the \lstinline{U}ser is stored in its
local variable \lstinline|address|. Since an AIOCJ program describes data
located at different participants, it uses the \lstinline|@|-notation to
indicate which participant owns a given variable --- in the case of the address,
it is \lstinline|address@U|. Then, we find the first interaction between
participants of the choreography, where the \lstinline{U}ser sends the address
to the \lstinline{W}ebUI. Following standard practice of
choreography models~\cite{survey}, all AIOCJ interactions are labelled --- in
this case, the label is \lstinline{getPage}. The notation %
\lstinline{U( address ) -> W( address )} means that \lstinline{U} sends the
value contained in its local variable \lstinline{address} to \lstinline{W},
which stores it in its local variable with the same name. Note the semicolon
\lstinline{;} between the first and second instructions. That symbol
represents a sequential composition operator that specifies that the instruction
on its left must execute before the instruction on its right. An AIOCJ program
describes a distributed system. Hence, to ensure the exact execution of
sequential compositions, the AIOCJ compiler makes sure that there is at least
one participant in common between two instructions composed in sequence ---
e.g., \lstinline|U| is the common participant between the two first instructions
--- to guarantee the faithful implementation of the causal relation among the
global actions in the choreography, which becomes distributed once compiled into
independent executables, communicating only via message passing. Then, we open a
scope at lines \ref{lst:bb-par-start} and \ref{lst:bb-par-end}, so that we
implement a fork-join pattern for parallelising the interaction between the
\lstinline{W}ebUI and resp.\@ the \lstinline{P}ersistence and the
\lstinline{I}mages services. We implement the forking part of the pattern with
two internal scopes (resp.\@ at lines
\ref{lst:bb-per-start}--\ref{lst:bb-per-end} and
\ref{lst:bb-img-start}--\ref{lst:bb-img-end}) that we compose using the parallel
composition operator \lstinline{|}. These two scopes execute in parallel and,
once terminated, join the larger sequential composition at the closure of the
wrapping scope at line \ref{lst:bb-par-end}. In the internal scopes, the
\lstinline{W}ebUI separately interacts with the \lstinline{P}ersistence and the
\lstinline{I}mages services to obtain the textual content of the page, saved in
the variable \lstinline{info} of the \lstinline{W}ebUI, and the images of the
page, saved in the \lstinline{img} variable of the \lstinline{W}ebUI. In each
scope, we resp.\@ find the \lstinline{P}ersistence and \lstinline{I}mages
participants that interact with the available APIs of the TeaStore services (or
wrappers thereof), resp\@. \lstinline{getPageInfo} and \lstinline{getPageImage},
to obtain the contents of the page (the name of the service API and the
homonymous label for the interaction are distinct concepts). In AIOCJ,
developers can introduce the availability of APIs through the
\lstinline{include} instruction, found at the beginning of AIOCJ programs, e.g.,
in \cref{lst:bb}, at lines
\ref{lst:bb-includes-start}--\ref{lst:bb-includes-end} --- where one can specify
the address and communication medium of the service (e.g.,
\lstinline{"socket://..."}) and the data format (e.g., \lstinline{"soap"}).

After the join, the \lstinline{W}ebUI continues by calling the
\lstinline{compilePage} API to obtain the \lstinline{page}, which it sends to
the \lstinline{U}ser.

Closing the example, we notice the presence of the \lstinline{preamble} clause
at line \ref{lst:bb-pre}, which developers can use to specify configuration
parameters, like the addresses of the participants. In this minimal example, we
let that AIOCJ compiler assign local addresses (for a sample local execution;
but AIOCJ supports general distributed executions), and we just indicate which,
among the choreography's participants, is the \lstinline{starter} --- this
information is necessary to implement a rendezvous procedure, where a
\lstinline{starter} participant, in this case, the \lstinline{W}ebUI, waits for
all the other participants to contact it and then notifies them they can start
the choreography.

\paragraph{Structuring Adaptation}
We close this section by showing how one can introduce structured, distributed
adaptation into AIOCJ programs. Specifically, we consider the case where other
APIs for the compilation of the page could become available in the future. 

Thanks to \lstinline{scope}s\footnote{AIOCJ supports two types of scopes:
generic scopes using the \lstinline|\{ ... \}| syntax and adaptation scopes
using the \lstinline|scope \{ ... \}| syntax. Throughout the paper, we use
syntax highlighting to distinguish between these forms in our examples. The
keyword \lstinline|scope| appears highlighted when referring to adaptable
scopes, while we use the word ``scope'' in regular text when referring to
generic scopes.}, AIOCJ allows developers to adapt the behaviour of the
participants in a choreography at runtime by integrating code written even after
they started their execution.

\begin{lstfloat}[t]
\begin{lstlisting}[language=MyChor,basicstyle=\ttfamily\footnotesize,
  caption={Excerpt of AIOCJ Choreography of an adaptable version of Barebone TeaStore.},
  label={lst:scope-compiler}]
scope @W {
 page@W = compilePage( info, img )
} prop { N.tag = "page-compiler" };
\end{lstlisting}
\end{lstfloat}

To illustrate this feature, we have to change the choreography from
\cref{lst:bb} by replacing line \ref{lst:bb-compile-page} with the code in
\cref{lst:scope-compiler}. In general, an adaptation \lstinline{scope} has two
elements: a controller, which is indicated with the \lstinline{@}-notation next
to the \lstinline{scope} --- in \cref{lst:scope-compiler}, the controller is the
\lstinline{W}ebUI --- and the body of the \lstinline{scope}, contained within
curly brackets, which delimits a piece of the choreography can change at runtime
through the application of AIOCJ \lstinline{rule}s --- discussed in the next
sections. 

At runtime, the controller is the participant in charge of selecting which
\lstinline{rule} applies to the \lstinline{scope}. While this example is simple,
i.e., the only participant in the \lstinline{scope} is also the controller, a
\lstinline{scope}
may involve
multiple participants, which the controller coordinates to make sure they all
follow the same piece of choreography found in a selected adaptation
\lstinline{rule}.
When a coordinator reaches the beginning of a \lstinline{scope}, it queries AIOCJ's runtime
for adaptation rules to apply. AIOCJ's runtime includes \emph{rule
repositories}, each containing one or more adaptation rules, which may (dis)connect at any time. The runtime queries the available rule repositories
sequentially which, in turn, check the applicability condition of each of their
rules. The runtime applies the first rule whose applicability condition holds,
if any, by sending to the coordinator the code for each participant, which it
distributes to the involved roles. In each role, the new code replaces the
original one. If no rule applies, the coordinator tells to the other roles to
execute the original code. This protocol ensures the consistency of adaptations,
which derive from the fact that there is a single source of ``truth'' that
determines the unfolding of the adaptation (the coordinator). Indeed, since
rules can change at runtime (they can appear and disappear), having only the
coordinator observe which rules are available solves the inherent problem of
inconsistencies (roles could see different sets of rules at different times and
locations) about rule availability and distribution of the related adaptation
code.

Developers can associate \lstinline{prop}erties with \lstinline{scope}s to make
the application of adaptation rules more precise, e.g., in
\cref{lst:scope-compiler}, we add the property \lstinline{tag}, prefixed by
\lstinline{N}\footnote{Originally, the prefix stood for ``non-functional'',
meaning that \lstinline{scope}-level properties should capture non-functional properties.
However, in practice, adaptation decisions often depend on interwoven functional
and non-functional concerns.}\footnote{The prefixing of \lstinline{scope}-level
properties/variables is not necessary, since these are syntactically and
contextually distinct from other variables (e.g., of the \lstinline{scope} controller). On
the contrary, the prefix is fundamental in rules, which can also consider other
kinds of properties in applicability conditions. For symmetry, AIOCJ imposes the
usage of the prefix also in \lstinline{scope}s.}, that labels the \lstinline{scope} as
carrying the code for the \lstinline{"page-compiler"}. Scope properties are
meant to describe the current implementation of the scope, including both
functional and non-functional properties. Such properties are declared by the
programmer, and the system only uses them to evaluate the applicability
condition of adaptation rules, to decide whether a given adaptation rule can be
applied to a given scope.

\begin{figure}[t]
  \centering
  \includegraphics[width=0.9\textwidth]{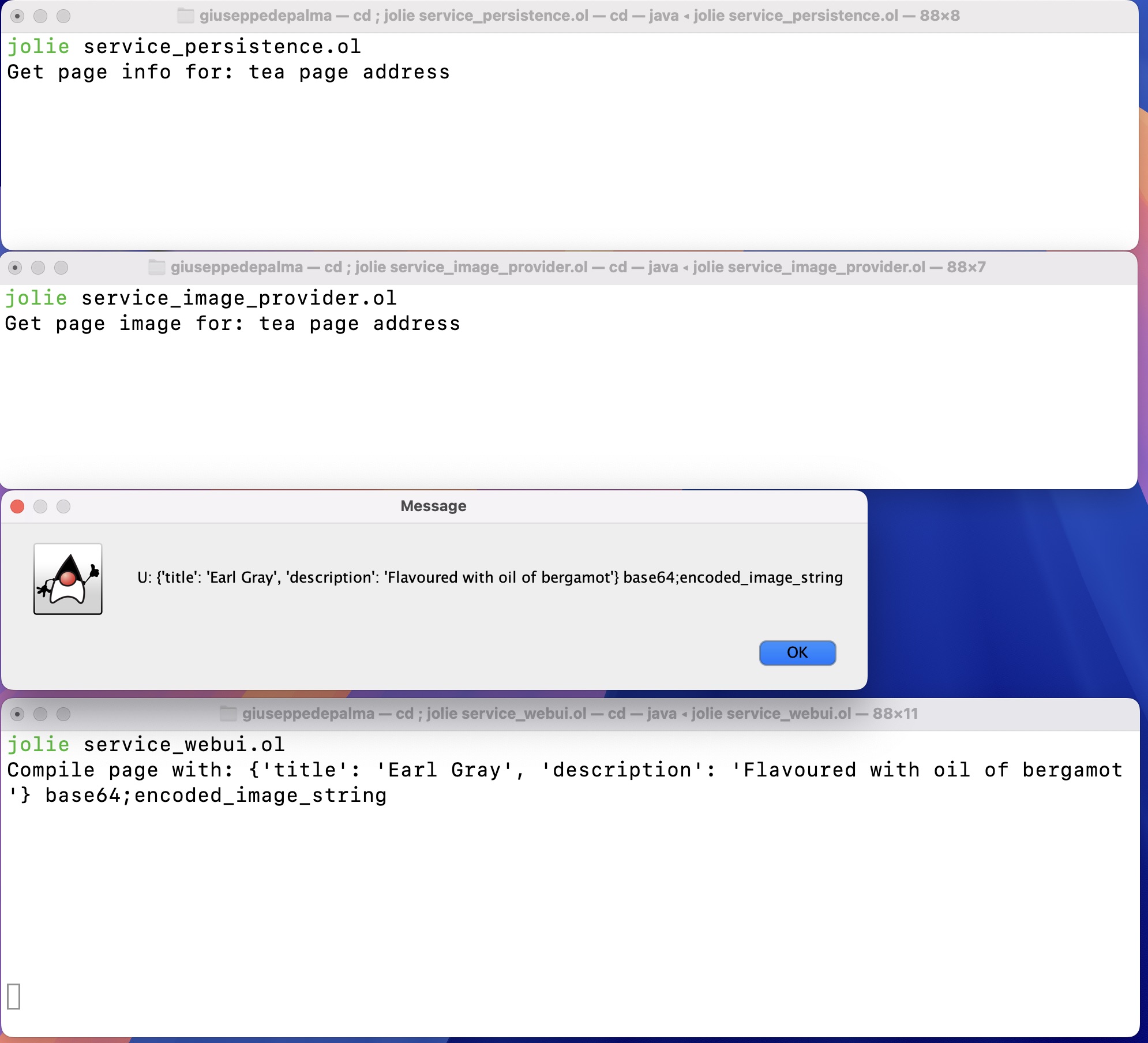}
  \caption{Execution of the Barebone TeaStore choreography.}
  \label{fig:screenshot-barebone}
\end{figure}

We further concretise our description by reporting in  
\Cref{fig:screenshot-barebone} a screenshot of the (local) execution of the
Barebone TeaStore choreography.

In the figure, we find a multi-terminal setup that demonstrates the execution of
services built with AIOCJ. Specifically, we show the behaviour of
Barebone TeaStore after the user inputs an address.

In \Cref{fig:screenshot-barebone}, a pop-up Message window appears showing the
result of the user interaction: a tea product with the title ``Earl Gray'', its
description, and an associated base64 encoded image. This output results from
the service invocations,
to the Persistence and Image Provider, found within the choreography.

Each terminal window runs
an independent Jolie~\cite{MGZ14} service (launched via the ``jolie'' command,
which executes the interpreter of Jolie programs). We can see two kinds of
services:
those that emulate/wrap the behaviour of the existing Adaptable TeaStore
services (e.g., the Persistence service, the top-most window),
and services automatically generated by AIOCJ's compiler. The latter include
both the services that implement the logic of the choreography participants
(omitted in \Cref{fig:screenshot-barebone}) and the services that constitute the
(distributed) runtime environment for AIOCJ's applications (e.g., an
\lstinline{Environment} service that lets administrators specify parameters of
the application execution context, also omitted in
\Cref{fig:screenshot-barebone}).

After inputting the address, the participants interact both among themselves and
with the external services (through their APIs) to retrieve the content of the
page requested, compile it, and show the result to the user.

\paragraph{Using Scopes to Implement Adaptation}
Summarising, AIOCJ provides \lstinline{scope}s as a way to specify which parts
of the application can be adapted. As a consequence, deciding which parts of the
code to enclose in \lstinline{scope}s is a relevant and non-trivial decision.
Intuitively, one should enclose into \lstinline{scope}s parts of the code that
may need to be adapted in the future. These parts include, e.g., the code
modelling business rules that may need to change according to changed business
needs, but also parts which are location dependent and parts which are relevant
for performance or security reasons. 

Of course, one could cut this Gordian Knot by having either a \lstinline{scope}
that encloses the whole choreography or many \lstinline{scope}s covering all
instructions. However, both solutions have relevant drawbacks. An
all-encompassing \lstinline{scope} would replace the whole choreography, which
means that one has to treat adaptation as a monolithic change that has to
integrate all relevant adaptation aspects (e.g., coalescing orthogonal
functional and non-functional aspects, like security and availability), giving
little-to-no support for modularity. Moreover, such an adaptation could only be
performed when the choreography starts. Using per-instruction scopes is hardly a
solution too, since a single adaptation
may involve
more than one scope and, at the moment, AIOCJ does not provide built-in ways to
structure multi-\lstinline{scope} adaptation behaviours --- as discussed in the
next sections, Adaptable TeaStore requires one such kind of coordination, which
we implicitly implement via bookkeeping.

\section{Barebone TeaStore Choreography with Adaptable Recommender}
As mentioned, Adaptable TeaStore allows for an optional Recommender, which comes
in two flavours: low-power, based on item popularity, and full-power, using
machine learning and user preferences.

To support such an adaptation
(like any other adaptation)
in AIOCJ
we need two
ingredients, a \lstinline{scope} specifying, in the code, where adaptation should
happen, and one or more adaptation \lstinline{rule}s, specifying which new code
should be used in case adaptation happens. Notably, adaptation rules can change
(be provided, removed) at runtime, during the execution of the original
choreography.

In this specific case, we use two \lstinline{scope}s. The first one, introduced
in the previous section, about page compilation, which allows one to exploit
information from the Recommender to produce a page for the user, including
recommendations, and one, shown in \Cref{lst:rec}, line~\ref{lst:scoperec}, to
insert the actual invocation to the Recommender -- the instruction
\lstinline|skip| denotes inaction. Note that this \lstinline|scope| introduces
an additional keyword, \lstinline|roles|, that indicates the possible
participation in the adapted code of roles that do not appear (i.e., do
something) in the body of the \lstinline{scope}. Specifically, the addition of
the \lstinline|U|ser and \lstinline|P|ersistence to the \lstinline|scope|
\lstinline|roles| allows adaptation rules to specify interactions involving
them, besides \lstinline|W|ebUI, found in the body of the \lstinline|scope|.

Note also that we put this new \lstinline{scope} in parallel
(\lstinline{|}) with the computation of the page information and the page image,
as discussed in the previous section.
An alternative would be to use just the \lstinline{scope} for page compilation
described in the previous section, with a rule introducing the Recommender as
well. While conceptually simpler, this alternative would be less efficient since
recommendation computation would start only after
the ones
of
page and image information have terminated.

\begin{lstfloat}[t]
\begin{lstlisting}[language=MyChor,basicstyle=\ttfamily\footnotesize,caption={Adaptable TeaStore barebone version, Recommender only},label={lst:rec}]
{
  {
    getPageInfo: W( address ) -> P( address );
    info@P = getPageInfo( pid );
    getInfo: P( info ) -> W( info )
  }
  |
  {
    getPageImg: W( address ) -> I( address );
    img@I = getPageImg( address );
    getImg: I( img ) -> W( img )
  }
  |
  $\label{lst:scoperec}$ scope @W { skip } prop { N.tag = "recommender" } roles { P, U }
};
\end{lstlisting}
\end{lstfloat}

We now describe the two adaptation rules, reported respectively in
\Cref{rule:low-power} and \Cref{rule:compilation}, starting from the former.

Adaptation rules have a main building block identified by the \lstinline{do}
clause, which specifies the code that needs to be executed in case adaptation is
performed, replacing the code inside the \lstinline{scope} under adaptation.
However, rules include other elements as well. First, they may include
additional external services (e.g., \lstinline{getTopItems}) and additional
participants, the latter introduced by keyword \lstinline{newRoles}. Finally,
the keyword \lstinline{on} introduces the applicability condition of the rule:
when a \lstinline{scope} is met during execution, all available rules are
checked for applicability (in no specific order). The first one whose condition
evaluates to true is applied. Conditions may refer to properties described by
the \lstinline{scope}, prefixed by \lstinline{N}, as in %
\lstinline{N.tag == "recommender"}, and to properties described by the
environment, prefixed by \lstinline{E}, as in %
\lstinline{E.recommender == "low-power"}. The former are meant to ensure that a
rule is applied to the ``correct'' \lstinline{scope} (i.e., the one relevant for
the mentioned property or functionality to be added),
while the latter allow rules applicability to depend on environmental condition,
e.g., if we are in an environment with limited power availability, so that the
rule in \cref{rule:low-power} applies.

\begin{lstfloat}[t]
\begin{lstlisting}[language=MyChor,basicstyle=\ttfamily\footnotesize,caption={Rule for the \textit{low-power} flavour of the Recommender service.}, label={rule:low-power}]
rule {
  include getTopItems from "socket://localhost:8001" with "soap"
  include processRecommendations from "socket://localhost:8003" with "soap"
  
  newRoles: R
  on { N.tag == "recommender" and E.recommender == "low-power" }
  do {
      getPopularProducts: R() -> P();
      items@P = getTopItems( 10, "popularity" );
      popularProducts: P( items ) -> R( items );
      recommendations@R = processRecommendations( items );
      recommendedProducts: R( recommendations ) -> W( recommendations );
      recommender@W = true
  }
}
\end{lstlisting}
\end{lstfloat}

The rule in \cref{rule:low-power} includes two external services that provide
the functionalities \lstinline{getTopItems} and \lstinline{processRecommendations}, respectively provided at the locations \lstinline{"socket://localhost:8001"} and \lstinline{"socket://localhost:8003"}, both using the \lstinline{soap} protocol.
Moreover, the rule introduces the new role \lstinline{R} (acting as coordinator
for the Recommender service) to the choreography. The rule activates when two
conditions are simultaneously met: the \lstinline{scope}'s \lstinline{tag}
equals \lstinline{"recommender"} and the environment variable
\lstinline{recommender} is set to \lstinline{"low-power"}. When triggered, the
rule implements a low-power recommendation flow where the
\lstinline{R}ecommender requests popular products from the
\lstinline{P}ersistence service, which then calls the \lstinline{getTopItems}
function to get the most popular items. \lstinline{P}ersistence sends these
popular items back to the \lstinline{R}ecommender, which processes them using
the \lstinline{processRecommendations} function. The \lstinline{R}ecommender
then sends the processed recommendations to the WebUI service, which sets its
\lstinline{recommender} flag to \lstinline{true}, indicating that the
recommendations are available.

While the previous rule allows one to compute recommendation information in
low-power mode, a second rule is needed to use this information at page
compilation. The rule is described in \Cref{rule:compilation}. Notably, the rule
showcases a third option one can use in applicability conditions: local
variables of the role in charge of managing the adaptation, as in
\lstinline{recommender == true}, where \lstinline{recommender} is a variable of
role \lstinline{W} (webUI). Here, \lstinline{recommender} is used as a bookkeeping variable, to ensure that adaptation is performed only if one of the rules in \cref{rule:low-power} has been applied. This mechanism allows one to synchronise different adaptation rules, albeit in an ad-hoc way.

\begin{lstfloat}[t]
\begin{lstlisting}[language=MyChor,basicstyle=\ttfamily\footnotesize,
  caption={Rule for the compilation of the page with recommendations.}, label={rule:compilation}]
rule {
  include compilePageWithRecommends from "socket://localhost:8000" with "soap"
  
  on { N.tag == "page-compiler" and recommender == true }
  do {
    page@W = compilePageWithRecommends( info, img, recommendations );
    recommender@W = false
  }
}
\end{lstlisting}
\end{lstfloat}

\begin{figure}[t]
  \centering
  \includegraphics[width=\textwidth]{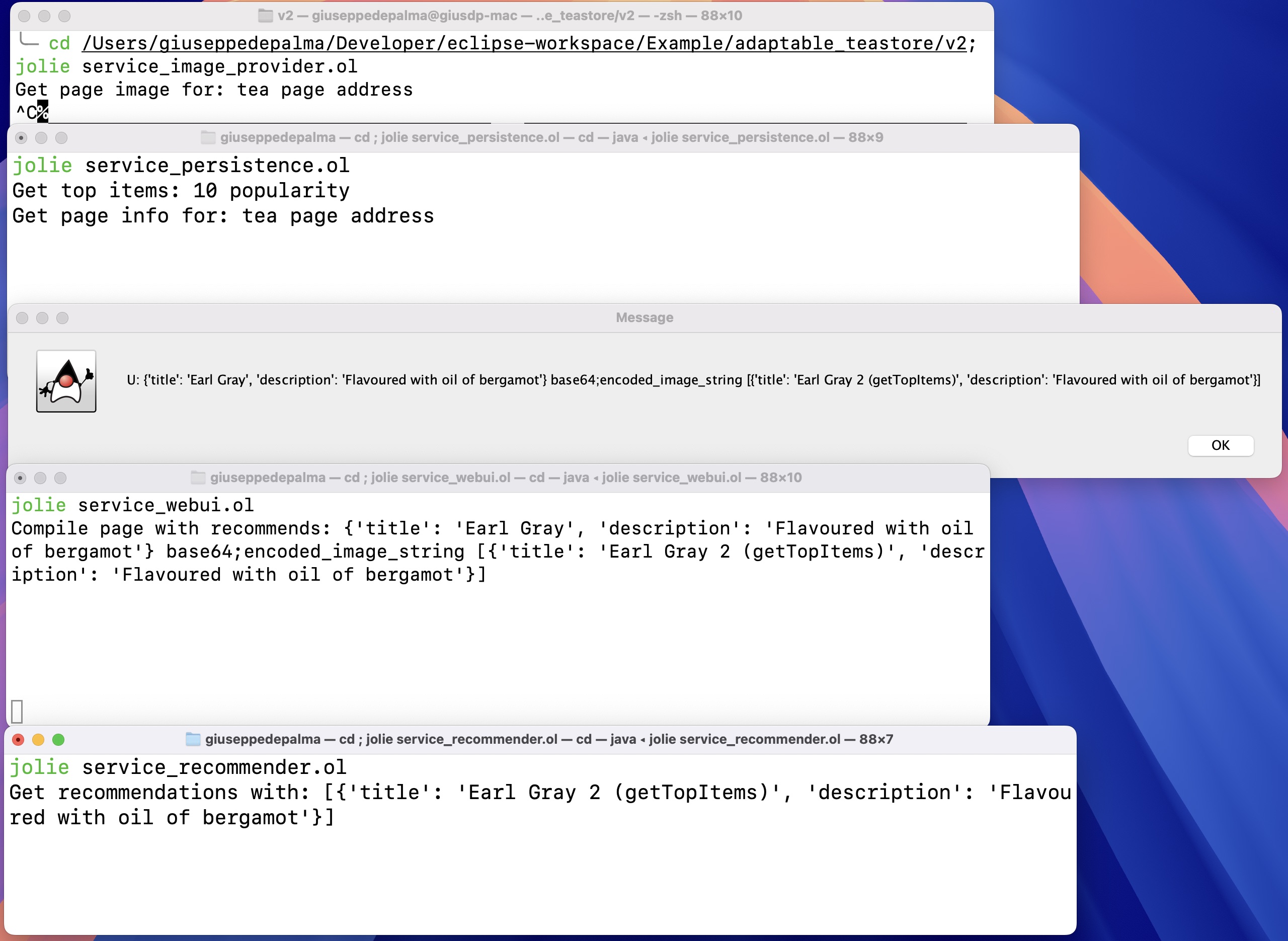}
  \caption{Execution of Barebone Adaptable TeaStore with the low-power Recommender version.}
  \label{fig:screenshot-recommender-low}
\end{figure}

As done in the previous section, we illustrate the execution of the adaptable
choreography with the screenshot reported in
\Cref{fig:screenshot-recommender-low}, which captures an execution step after
the application of the rule for the adaptation with the low-power version of the
Recommender. In particular, we notice that the resulting message to the user is similar to the one in \Cref{fig:screenshot-barebone}, although, in \Cref{fig:screenshot-recommender-low}, it integrates the suggestion of the Recommender.

\paragraph{On the Application of Adaptation Rules}
When, during execution, the runtime enters a \lstinline{scope}, it checks all available
rules for applicability (in no specific order) and it applies the first one
whose condition evaluates to true. Specifically, the runtime performs the check
for applicability in sequence, considering the rules one by one w.r.t.\@ the
current status of the system. Since environmental properties can change in
parallel with rule evaluation, race conditions can occur where the applicability
condition of a rule is negative at check time but becomes positive afterwards,
leading to scenarios where the runtime applies no rule, even despite having
complete coverage of all possible states\footnote{Consider rules that apply
w.r.t.\@ the state of a database connection, which can either be ``primary'',
``replica'' or ``offline''. Let us assume that we enter an adaptation scope, the
current status is ``primary'', and that the runtime proceeds to check the
applicability of a rule that requires the state to be ``replica''. The condition
is negative, and the runtime moves on to check the next rules. Meanwhile, the
state changes to ``replica'', which prevents the application of the other two
rules. The result is that no rule applies, even though the system has complete coverage for all the database states.}.

This issue is not easily amendable due to the concurrent nature of distributed
architectures. Attempting to fix the environment state during rule checking
(e.g., by taking a snapshot or locking environmental variables) would be
counterproductive, as it would cease to provide a valid representation of the
actual system status and could lead to applying rules based on desynchronised
states. Moreover, the possibility that the runtime applies a rule while the
system status changes (potentially causing errors) is not fundamentally
different from other distributed system failure modes. For instance, a
choreography might successfully adapt and then have an external resource crash
during execution, leading to similar error conditions. We argue that one might
more effectively address these race conditions through error handling and
recovery mechanisms rather than attempting to eliminate the temporal
inconsistencies that are inherent to distributed systems. The challenge lies in
designing adaptation strategies that are resilient to these timing issues. We refer the reader interested in a thorough description of the theory and implementation of AIOCJ's adaptation mechanisms and guarantees to its reference presentation~\cite{LMCS2017-dynamicChoreo}.

\section{Barebone TeaStore Choreography, with Adaptable Authentication}
We now consider a more complex adaptation scenario, where the Authentication
service is added. The idea is that, if the user can authenticate/is
authenticated, the system can provide them with personalised products, images,
and recommendations, based on their preferences.

However, we have a tricky point on our hands: authentication impacts the
choreography at many points, including when services, such as the Recommender,
are possibly added at runtime. To coordinate and follow a consistent behaviour,
we exploit a variable \lstinline{token}, managed by the WebUI, to keep track of
whether authentication has already been performed or not (cf.~\Cref{lst:token},
line~\ref{lst:tokenline}). We also add various \lstinline{scope}s to enable adaptation.

\begin{lstfloat}[t]
\begin{lstlisting}[language=MyChor,basicstyle=\ttfamily\footnotesize,caption={Adaptable TeaStore barebone version},label={lst:token}]
aioc {
  address@U = getInput( "Insert address" ); 
  getPage: U( address ) -> W( address );
  $\label{lst:tokenline}$token@W = "none";
  {
    scope @W {
      getPageInfo: W( address ) -> P( address );
      info@P = getPageInfo( address );
      getInfo: P( info ) -> W( info )
    } prop { N.tag = "page-info" } roles { U }
    |
    scope @W {
      getPageImg: W( address ) -> I( address );
      img@I = getPageImg( address );
      getImg: I( img ) -> W( img )
    } prop { N.tag = "page-images" } roles { U }
    |
    scope @W { skip } prop { N.tag = "recommender" } roles { U, P }
  };
  scope @W {
    page@W = compilePage( info, img )
  } prop { N.tag = "page-compiler" };
  getPage: W( page ) -> U( page );
}
\end{lstlisting}
\end{lstfloat}

The main rule providing authentication is described in~\Cref{lst:auth}. Its
applicability condition specifies that it applies to \lstinline{scope}s  \lstinline{"auth"}
and requires authentication support to be available %
(by checking \lstinline{E.auth == "available"}). Note that the rule applies only if no token
has been obtained yet. These conditions model the fact that authentication
should be ideally performed once, providing a token that can be used by all the
functionalities needing it.

\begin{lstfloat}[t]
\begin{lstlisting}[language=MyChor,basicstyle=\ttfamily\footnotesize,caption={Rule for the Authentication service.},label={lst:auth}]
rule {
  include login from "socket://localhost:8004" with "soap"
  newRoles: A
  on { N.tag == "auth" and E.auth == "available" and token == "none" }
  do {
    credentials@U = getInput( "Insert Credentials" );
    sendCredentials: U( credentials ) -> A( credentials );
    token@A = login( credentials );
    if ( token != "none" )@A {
      sendToken: A( token ) -> W( token )
    }
  }
}
\end{lstlisting}
\end{lstfloat}

\Cref{lst:prodInfo} shows the rule for refined page info compilation. In
practice, the service which requires authentication
\lstinline{getPageInfoAsLoggedUser} is used if either the user authenticated
beforehand and the system has
their
token, or if the scope inside the rule at
line~\ref{lst:authline} is updated, thus obtaining the token ``on the fly''.

The rule in \Cref{lst:prodInfo} shows that we can have nested adaptations by
using scopes inside rules. In this case, once the rule is applied and the
execution reaches the nested \lstinline|scope| at line \cref{lst:authline} of
\Cref{lst:prodInfo}, the system will again check for the availability of
adaptation rules (specifically, with \lstinline{N.tag = "auth"}). Note that the
rule in \Cref{lst:prodInfo} is applied independently of the availability of
authentication facilities (e.g., there is no condition %
\lstinline{E.auth == "available"}). This is meaningful in a scenario where
authentication facilities may appear and disappear during the computation, e.g.,
the user authenticated at a previous stage, the authentication facility is not
available any more, but we can still use the token to process the user's
requests.

\begin{lstfloat}[t]
\begin{lstlisting}[language=MyChor,basicstyle=\ttfamily\footnotesize,caption={Rule for the Persistent interaction to retrieve product info.},label={lst:prodInfo}]
rule {
  include getPageInfo, getPageInfoAsLoggedUser from "socket://localhost:8001" 
    with "soap"
  on { N.tag == "page-info" }
  do {
    getPageInfo: W( address ) -> P( address );
    $\label{lst:authline}$scope @W { skip } prop { N.tag = "auth" } roles { U };
    if ( token != "none" )@W {
      sendToken: W( token ) -> P( token );
      info@P = getPageInfoAsLoggedUser( address, token )
    } else {
      info@P = getPageInfo( address )
    };
    getInfo: P( info ) -> W( info )
  }
}
\end{lstlisting}
\end{lstfloat}

\Cref{lst:full-power} provides adaptation for the full-power version of the
Recommender. The logic of this rule is similar to the one from
\cref{lst:prodInfo}, where nested adaptation provides authentication when
needed.

\begin{lstfloat}[t]
\begin{lstlisting}[language=MyChor,basicstyle=\ttfamily\footnotesize,
  caption={Rule for the \textit{full-power} flavour of the Recommender service.},label={lst:full-power}]
rule {
  include getPageInfo, processQuery, 
          getPageInfoAsLoggedUser from "socket://localhost:8001" with "soap"
  include getQuery, getQueryAsLoggedUser, 
          processRecommendations from "socket://localhost:8003" with "soap"
  newRoles: R
  on { 
    N.tag == "recommender" and E.recommender == "full-power" 
  }
  do {
    getPageInfo: W( address ) -> P( address );
    scope @W { skip } prop { N.tag = "auth" } roles { U };
    if ( token != "none" )@W {
      sendToken: W( token ) -> P( token );
      info@P = getPageInfoAsLoggedUser( address, token );
      getInfo: P( info ) -> R( info );
      sendToken: W( token ) -> R( token );
      query@R = getQueryAsLoggedUser( info, token )
    } else {
      info@P = getPageInfo( address );
      getInfo: P( info ) -> R( info );
      query@R = getQuery( info )
    };
    sendQuery: R( query ) -> P( query );
    result@P = processQuery( query );
    queryResult: P( result ) -> R( result );
    recommendations@R = processRecommendations( result ); 
    recommendedProducts: R( recommendations ) -> W( recommendations ); 
    recommender@W = true 
  }
}
\end{lstlisting}
\end{lstfloat}

Note that the \lstinline{scope}s which can provide adaptation are in
parallel in \Cref{lst:token}. Hence, depending on the scheduling, multiple
authentications may be performed. This behaviour happens, in particular, if the
conditions for applicability of the authorisation rule are all checked before
the variable token gets a \lstinline{"none"} value. On the one hand, apart for
the burden for the user to authenticate multiple times, no issues is caused. On
the other hand, avoiding this would require to sequentialise the different
steps, losing efficiency.

\begin{figure}[t]
  \centering
  \includegraphics[width=\textwidth]{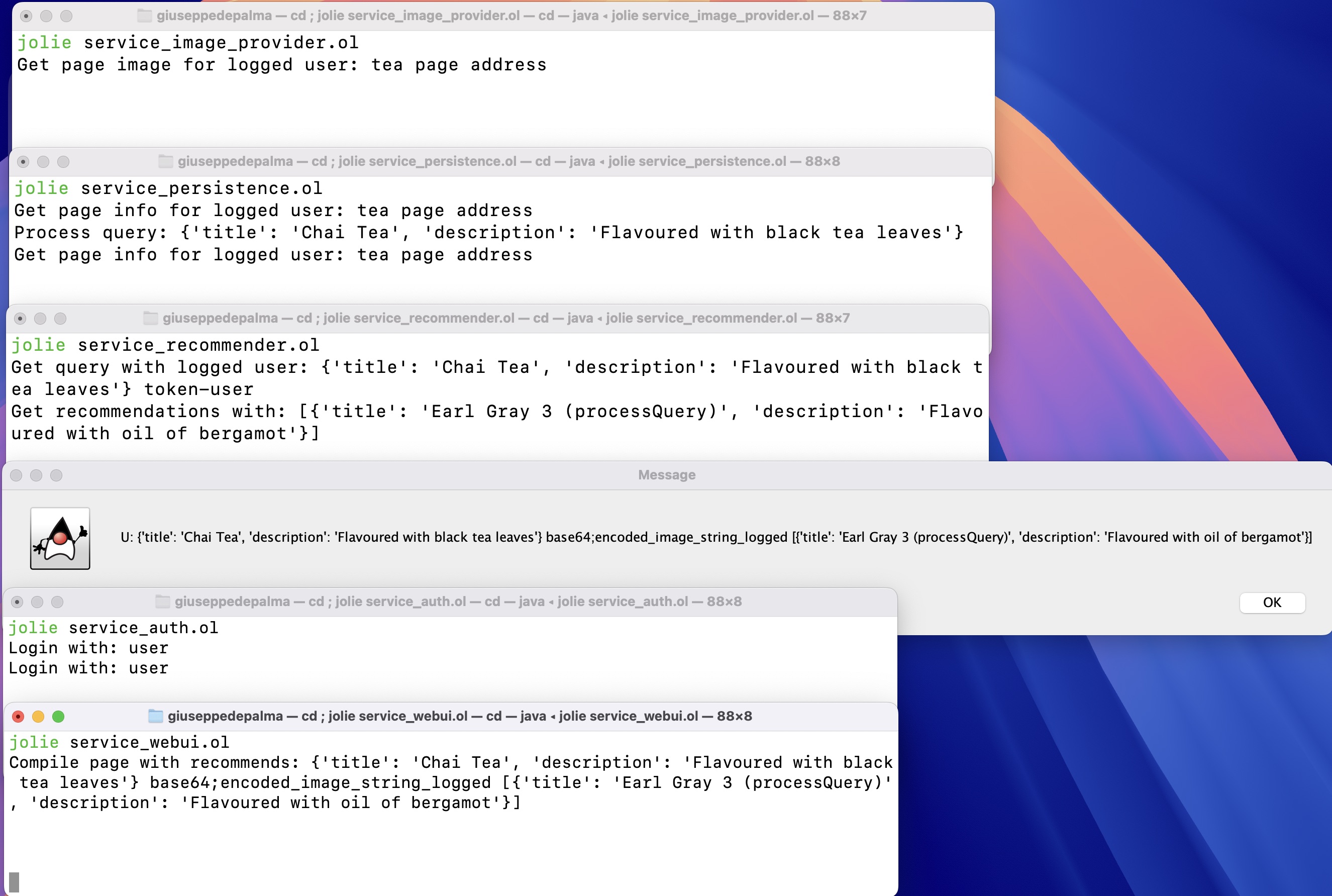}
  \caption{\label{fig:screenshot-recommender-full}Execution of the Adaptable
  TeaStore choreography with the Auth service and the full-power Recommender
  version.}
\end{figure}

Closing the section, we report in \Cref{fig:screenshot-recommender-full} a
screenshot of the execution of the Adaptable TeaStore choreography after the
full adaptation using the Auth service and the full-power Recommender. In the
figure, we can see that  the Auth service allowed the user to log in (visibile
in the Auth service console; the login happens two times due to the parallel
execution of the \lstinline{scope}s in \cref{lst:token}). We also notice the
execution of the full-power version of the Recommender from the result of the
page compilation in the pop-up window, where we find the recommendation with
`description': `Flavoured with \dots'.

\section{Discussion and Conclusion}

We see this contribution as a useful benchmark of the fitness of a language such
as AIOCJ (and, to some extent, the choreographic programming approach) in
capturing the concerns and possibilities of the TeaStore architecture and
helpful to indicate and discuss implementation patterns and styles inspired by
the choreographic paradigm.

\paragraph{Ephemeral Adaptations and Implicit Regressions}
Our approach with AIOCJ starts from a barebone system, where we strategically
insert adaptation \lstinline{scope}s at points where we anticipate potential
adaptation needs. The application of adaptation rules relative to the available
\lstinline{scope}s and system environment states then adapts the system using
the appropriate choreography fragments (found in the rules) to replace
these \lstinline{scope}s. This methodology implies that, for each interaction
flow among the participants (i.e., the execution of the base choreography), the
system adapts according to available rules and current conditions.

A significant advantage of this approach is that we do not need to specify
``regression'' adaptation rules, because the system always starts executing
from the original version and possibly adapts according to the available rules
and rule applicability conditions. This point becomes more practical if we
consider ``wrapping'' the behaviour found within the \lstinline{aioc} scope in
the examples with a \lstinline|while(e)@W{ ... }|\footnote{From the language standpoint, similarly to \lstinline|if|s,
\lstinline|while|s require the indication of a controller, i.e., a participant
(in the example above, \lstinline|W|) that determines whether the participants
of the loop (including the controller) shall execute the code within the
\lstinline|while| scope or continue with the next instruction.}, making the whole system loop
(as long as the evaluation of expression \lstinline|e| at \lstinline|W|
evaluates to true) and implementing in a more faithful way the behaviour of the
Adaptable TeaStore --- where the User can interact with (and request content
from) the WebUI multiple times. Considering the full Adaptable TeaStore
implementation (from \cref{lst:token}), using the \lstinline|while| loop would
mean that, at each iteration, the system would start from the ``barebone''
version and then adapt according to the current status of the system --- e.g.,
during an iteration the Auth service might be unavailable, preventing the
application of (parts of the) rules that concern the latter, while at the next
one the service might become available, supporting the application of the
related rules.

Generally, we can describe the application of AIOCJ adaptations as
``ephemeral'' (i.e., bound to a given flow iteration, but otherwise volatile
w.r.t.\@ the overall system status) which, complementarily, determines the
implicit regression of the system to the original state.

While the ephemeral application of adaptation provides high flexibility, we
notice that this pattern can introduce potential performance costs, since
adaptation can occur multiple times (e.g., at each iteration and, separately, at
different execution steps of the choreography), and ``fluctuations'' in user
experience and system states, if rules are not applied deterministically. For
instance, a user could initially successfully log in because the Auth service is
available; then, due to the non-deterministic application of rules, the rule
enabling the usage of the Auth service might not apply, and the user might
experience an ``inconsistent'' behaviour, given their previous successful login.
That said, in conventional systems, if the Auth service fails, users would
typically encounter a standard error (e.g., HTTP 500), determining fluctuations
similar to the one mentioned above. In general, we consider exploring the
practical implications of applying the ephemeral adaptation pattern in these
contexts an interesting future endeavour for refining the (approach behind the)
AIOCJ language.

\paragraph{Integrating Service Controllers}
One important element we abstracted away in our modelling is that, at the AIOCJ
level, we do not manage provisioning, scaling, or other infrastructure concerns.
AIOCJ, at the moment, provides no structured way to specify or reason about
these concerns, which we envision incorporating into choreographies in the
future.

In practice, one could integrate the AIOCJ language and runtime with external
controllers for service (de)allocation, such as Kubernetes. These controllers
can react to various events like traffic increases or node failures by
allocating services, relocating them, etc. The services they manage are the
Adaptable TeaStore components interconnected through our choreography. Thus, one
should have a way of integrating service controllers with choreographic
implementations so that the availability of services, load balancing, and
similar concerns are integrated within the choreographic approach.

\paragraph{Modelling Functional and Non-Functional Concerns in AIOCJ}

The discussion above, on integrating AIOCJ constructs with architecture
controllers, introduces a more general issue, which is the separation of
functional and non-functional concerns --- the latter include performance,
availability, scalability, resilience, and reliability. Specifically, while
AIOCJ does not explicitly model these aspects, they can surreptitiously show up
in adaptation code. Considering our examples, we adapt the architecture
according to the presence/absence of certain components, as witnessed by the
applicability conditions %
\lstinline|E.recommender == "low-power"| and \lstinline|E.auth == "available"|
found within the adaptation rules, which e.g., specify the handling of different
service flavours as part of the functional specification.

Hence, at the moment, AIOCJ provides no structured way to specify or reason
directly about these non-functional properties which, nonetheless, are relevant
to adaptation. This mixing surreptitiously introduces a hybrid approach where
some non-functional aspects are choreographically coordinated while others are
delegated to external controllers, like Kubernetes, which further strengthens
the point for integrating external controllers at the choreographic level.

\paragraph{Adaptation Compositional Complexity}

An alternative to the way we modelled the compositionality of Adaptable TeaStore
scenarios and configurations involves creating rules for each service
configuration, e.g., rules for ``Auth + Recommender full-power'', ``Auth +
Persistence'', etc. The challenge with this approach is the need for numerous
rules. Indeed, while this alternative pattern could, in principle, lead to more
easy-to-interpret adaptation scenarios (there is no need to ``figure out'' what
combination of rule applications one could obtain, depending on the availability
of \lstinline{scope}s and rules) it could lead to an ``explosion'' of adaptation rules,
depending on the coupling between participant/service behaviours within the
choreography. For this reason, in this paper, we opted for the nesting of rules,
e.g., as seen in \cref{lst:prodInfo} and \cref{lst:full-power}. Another
interesting research direction is to explore in which contexts one pattern might
be more suitable than the other, considering both qualitative traits, e.g., in
terms of how easily programmers can specify the expected behaviour of an
adaptable system, and in terms of performance, e.g., given that nested
adaptations may involve more runtime steps than the ``flat'' ones.

\paragraph{Choreographic Service Composition}

As a general note, we observe that Adaptable TeaStore (like TeaStore) is highly
orchestration-oriented, with WebUI orchestrating services provided by other
components. In this paper, we faithfully model TeaStore's behaviour,
implementing a choreography that follows an idiosyncratic orchestration-oriented
composition pattern (where the WebUI centralises most interactions), future
extensions of this work could propose alternative, idiomatic choreographic
versions of the TeaStore, e.g., where we support direct communication between
services without routing them through the WebUI, e.g., for increased efficiency.

We argue that the apparent mismatch between Adaptable TeaStore's architectural
pattern (centralised orchestration) and the AIOCJ modelling approach
(distributed choreography) helps us to demonstrate the flexibility of AIOCJ in handling
non-idiomatic patterns, showing how choreographic approaches can refactor
orchestration-oriented systems and that choreographic correctness guarantees
(like deadlock freedom) remain valuable even when modelling centralised
architectures.

\paragraph{Error handling}
While, for brevity, we did not discuss the issue of interacting with failing
services from an AIOCJ choreography, we notice that the language gives little
support to service interaction handling.

Indeed, when a choreography communicates with an external service, if the latter
fails, the entire choreography breaks down. This risk represents a
fundamental design challenge in the AIOCJ model.

To prevent stopping the execution due to an external service failure, the AIOCJ
runtime should include an intermediary service layer (that cannot fail and)
capable of handling requests from the choreography and returning values
indicating success or failure of the operation.
Such a mediator would enable adaptation rules to be triggered based on service
availability. For instance, when contacting the Auth service and receiving a
response indicating the service is unreachable, the system could apply an
appropriate adaptation rule specifically designed for this scenario. The current
model lacks this intermediary resilience layer, creating a brittle dependency
between choreographies and external services.

Alternatively, one might introduce choreography-level error handling mechanisms
that directly trigger adaptation when external service failures occur. This
extension would require that AIOCJ incorporates exception handling constructs
(such as the ones proposed in~\cite{CapecchiGY10}) that can seamlessly
transition into adaptation scenarios.

Generalising, the discussion above exposes a broader issue with choreographic
approaches like AIOCJ: while they excel at ensuring correctness properties
during normal operation and adaptation transitions, they struggle with graceful
degradation when facing unexpected external failures. The strong coupling
between choreographic descriptions and external services creates a single point
of failure that contradicts the resilience goals of modern Cloud
architectures.
Furthermore, this issue compounds when adaptation rules themselves depend on
potentially failing external services. In such cases, the very mechanism
designed to handle changing requirements becomes vulnerable to the same failure
modes it aims to address.

Considering the specific case of Adaptable TeaStore, the aspect of error
handling becomes particularly evident when considering the architecture's
multiple service variants and failover scenarios. While AIOCJ can express the
transitions between service configurations, it struggles to handle the detection
and management of the failures that would require such transitions in the first
place.

A strategy towards supporting error handling in AIOCJ is having monitors that
track the state of external services and feed this information to the
environment. When external services go down, the environment becomes aware of
this status change and can trigger appropriate adaptations. However, this
monitoring approach has inherent limitations. It only works effectively if the
monitor has an updated view of the status of the external service when the \lstinline{scope}
begins and when the adaptation check is performed. More critically, this
approach provides no protection if the external service breaks during the
execution of the \lstinline{scope}, after the adaptation check has already
occurred. This timing vulnerability creates a window of failure that cannot be
addressed through conventional adaptation mechanisms in AIOCJ, highlighting the
need for more robust error handling capabilities integrated directly into the
choreographic model.

\paragraph{Adaptation flows}
Another interesting refinement point for the AIOCJ language emerges from the
usage of bookkeeping variables to coordinate the behaviour of different,
adaptable parts of the choreography. Indeed, in our AIOCJ model of the Adaptable
TeaStore, we introduced several variables to track the state and availability of
services across adaptation scenarios. For instance, we needed variables to
record whether the Auth service was available and which version of the
Recommender was active. These variables are not part of the conceptual model of
the TeaStore but represent implementation artefacts required to bridge the gap
between AIOCJ's adaptation mechanisms and the actual system state.

The use of these bookkeeping variables introduces several problems, such as
making the choreography description more complex and harder to understand, as
readers must mentally track the state of these variables alongside the actual
business logic. Moreover, bookkeeping creates potential for inconsistencies, as
variables might not be properly updated in all execution paths, making also
proposing static verification techniques more difficult, as formal analysis must account
for these additional state elements.

To address these limitations, refinements to the AIOCJ language could introduce
direct support for adaptation based on service states (e.g., integrating runtime
interaction of the rule with orchestrators, such as Kubernetes), which would
make adaptation conditions more declarative and closely aligned with the
conceptual model of the system.

\paragraph{Other Approaches}
The two approaches closest to AIOCJ we are aware of are based on
multiparty session types~\cite{survey}. The first, by Anderson and Rathke, deals
with dynamic software updates~\cite{DSUtypes}. The second, by Coppo et al.,
regards monitoring of self-adaptive systems~\cite{coppo-multiparty-adaptation}.
The main difference between Anderson and Rathke's work~\cite{DSUtypes} and AIOCJ
is that the former targets concurrent applications which are not distributed.
Indeed, it relies on a check on the global state of the application to ensure
that the update is safe. Such a check cannot be performed by a single role, thus
it is impractical in a distributed setting.
Furthermore, the language by Anderson and Rathke~\cite{DSUtypes} is more
constrained than ours, e.g., requiring each pair of participants to interact on
a dedicated pair of channels, and assuming that all the roles that are not the
sender or the receiver within a choice behave the same in the two branches.
The approach by Coppo et al.~\cite{coppo-multiparty-adaptation} is also quite
different from ours. In particular, Coppo et
al.~\cite{coppo-multiparty-adaptation} require possible behaviours to be
available since the beginning, both at the level of types and of processes, and
a fixed adaptation function is used to switch between them. This difference
derives from the distinction between self-adaptive applications, as they
discuss, and applications updated from the outside, as in our case.

We also recall the work by Di Giusto and P\'erez~\cite{DP13}, who use types to
ensure safe adaptation. However, in that work, updates can happen only when no
session is active, while we change the behaviour of running choreographies. We
highlight that, contrarily to our approach, none of the approaches above has
been implemented.

More recently, Harvey et al.~\cite{HFDG21} presented EnsembleS, an actor-based
language that uses multiparty session types to provide static, compile-time
verification for safe runtime adaptation. Both AIOCJ and EnsembleS aim to
guarantee communication safety in adapting systems. EnsembleS ensures safety by
checking the compatibility of newly discovered components against predefined
session types at compile time. In contrast, AIOCJ's guarantees are
by-construction and allow for dynamic runtime changes that are defined by
adaptation rules rather than a static type-checking process.

Broadening our scope, we find proposals like the one by Herry et
al.~\cite{HAR13} and Wild et al.~\cite{WBKLW20}, who use choreographies for
automating the decentralised application of deployment. These approaches use a
global declarative deployment model that is split into local parts for each
participant. Workflows are then generated from these local models, forming a
deployment choreography that coordinates local deployments and
cross-organisational data exchange. In a similar direction, Philippe et
al.~\cite{POCPR24} presented Ballet, a decentralised choreography-based approach
for reconfiguring distributed systems. These approaches align with AIOCJ in
their use of choreographies to achieve decentralised control and coordination.
However, Herry et al.'s, Wild et al.'s, and Philippe et al.'s works focus on
deployment and coordination of application components. On the contrary, AIOCJ
captures ongoing, dynamic runtime adaptation of microservice architectures.
We deem interesting the study of how to integrate AIOCJ with this kind of works to extend the language/runtime's coverage to components deployment.

In another recent contribution, Ortiz et al.\@ presented an approach for
adaptation in microservice choreographies through event-based BPMN fragment
compositions~\cite{OTV25}. In that work, local changes to a microservice are
propagated using a catalogue of adaptation rules that aim to preserve the
functional integrity of the global process. However, since the catalogue is
finite, situations may arise where no suitable rule applies, leading to
inconsistencies or even failures in the choreography. This possibility contrasts
with AIOCJ's choreographic approach, where if no adaptation rule is available,
the system safely defaults to the original behaviour.

For a broader and thorough discussion on adaptation techniques for
distributed software systems, mainly focussed on component-based approaches, we
refer to the recent survey by Coullon et al.\cite{CHLR23}.

Finally, we mention Aspect-Oriented Programming (AOP)~\cite{KLMMLLI97} and
Context-Oriented Programming (COP)~\cite{HCN08}. Both programming paradigms aim
to improve code flexibility. AOP addresses cross-cutting concerns --
functionality that spans multiple software units, like logging, security, or
transaction management and, instead of scattering code throughout the codebase,
it gathers them into ``aspects'' that can be woven into the code at ``join
points''. COP focuses on making programs adapt their behaviour based on the
current execution context or environment. Rather than having fixed behaviour,
components can dynamically change how they operate depending on factors like
user preferences, system state, or external conditions. The application of these
approaches have been proposed at the Workshop on Adaptable Cloud
Architectures (WACA) 2025 by Truyen~\cite{T25} as a way to decouple the control
logic from applications, express fine-grained adaptation, determine system-wide
control, and reuse traditional adaptation strategies. In this context, AIOCJ can
deal with cross-cutting concerns like logging and authentication, typical of
AOP, by viewing ``pointcuts'' (where aspects should be applied) as empty scopes
and ``advices'' (what code to execute and when) as adaptation rules. Layers,
COP's modular units of behavioural variation that can be dynamically activated
or deactivated based on context, can instead be defined in AIOCJ by adaptation
rules which apply according to contextual conditions. Better understanding the relations between AIOCJ and AOP/COP is an interesting aim for future work.

\bibliographystyle{eptcs}
\bibliography{biblio}
\end{document}